**Fungal Genetic Variants in Oceanic Environments**

**Authors:** Sade A. Davenport[1] and Scott H. Harrison[2,*]

**Affiliations**
[1]Intelligence Community Postdoctoral Research Fellowship Program, North Carolina Agricultural and Technical State University, 1601 E Market St, Greensboro, NC 27411.
[2]Department of Biology, North Carolina A&T State University, 1601 E. Market St, Greensboro, NC 27411

*Corresponding author; scotth@ncat.edu

E-mail addresses
sadedavenport17@gmail.com (S. Davenport); scotth@ncat.edu (S. Harrison)

**Abstract**

Comparing specific types of organisms as they are found across environmental conditions has helped inform how genes and gene products of these organisms relate to phenotypes and adaptation. In this study, we examine metatranscriptomic data as found for oceanic fungi across different oceanic sampling sites. A specific set of three genes was chosen for evaluation based on conserved orthology, known association with core physiological processes in fungi, and level of abundance within oceanic metatranscriptomic data. We report upon a potential association of genetic variance with environmental conditions of iron, salt and phosphate in oceanic waters based on heatmap visualization and PERMANOVA analysis.

**Keywords**
fungi, cell cycle genes, ocean, metatranscriptomic, PERMANOVA

**1. Introduction**

Researching oceanic microorganisms has been in the past challenging due to expensive ship time and prior approaches involving the need to acquire samples aseptically, in conjunction with the measurement of environmental parameters associated with specific depths and water columns of sampling (Fell, 2012). Recent advances have led to initiatives such as the Tara Oceans project using metagenomic- and metatranscriptomic-based methods for studying the prevalence and diversity of oceanic microorganisms. The Tara Oceans project provides short read sequencing data of fungi and other microorganisms along with environmental metadata from different sampling sites across Earth's various oceans (Villar et al., 2018). This provides a powerful and novel opportunity to directly compare variation in genetic sequence from metagenomic and metatranscriptomic samplings across the broad-ranging oceanic environment. This can aid discovery with both patterns of species-level diversity across the ocean and how genetic features may differ in their association with respect to varying physical and chemical conditions across the oceanic environment.

Aquatic fungi such as aquatic yeast were initially reported in the 19th century and are widespread (Kutty & Philip, 2014; Braun, 1856). Marine yeasts are found in environments ranging from the oceans, salt lakes, deep sea sediments, and marshes. Early research on marine yeasts reported on the general occurrence and distribution of yeasts in the ocean. Subsequent research studies have charted the taxonomy of marine fungal diversity by characterizing new species and other more broad taxa, with attention given to these varieties' ecological roles within the marine environments. Major reports on marine fungi include those by Hagler (1987), Fell (2012), Kriss (1967) and Morris (1968). Our study is oriented upon how marine fungi in particular can be studied through the use of metatranscriptomic data from different ocean sampling locations. In this study, we used distance matrix (heatmap) visualizations and PERMANOVA analyses of abundance to identify the potential association of different fungal genetic sequences with different physical and chemical conditions of the oceanic environment.

## 2. Methodology

### 2.1. Collecting Representative Protein Sequences for Analysis

This study first gathered together protein sequences of kinases found for yeasts along with associated annotations and aspects of their features. Data for these yeast kinases were obtained using the Yeast Kinase and Phosphatase Interactome (KPI) Resource (https://yeastkinome.org/, retrieved 04/08/2021; Breitkreutz et al, 2010). This KPI database contains *Saccharomyces cerevisiae* kinase and phosphatase interactome data and was screened for those kinases that were: 1) annotated with direct evidence concerning phosphorylation, and 2) expected to be conserved across fungal diversity through involvement in the core physiological process of cell cycling. Protein sequences corresponding to each of the following five kinase genes from the genome of *S. cerevisiae* strain ATCC 204508 / S288c were as follows, listed by gene name and corresponding NCBI accession number: CBK1 (QHB11221.1), KIN1 (KZV12362.1), KIN2 (CAA97659.1), CDC28 (CAA85119.1), and DBF20 (KZV07624.1). These sequences were then retrieved from the NCBI Protein database (https://www.ncbi.nlm.nih.gov/protein, retrieved 04/08/2021) (Geer et al., 2010).

### 2.2. Metatranscriptomic Data

The MATOU database of the Tara Oceans project is a metatranscriptomic data collection that provides more than 116 million distinct genetic sequences that are assigned MATOU IDs (i.e., Marine Atlas of Tara Oceans Unigenes) (Hingamp, 2018). Each MATOU ID represents a distinct transcribed gene sequences. Gene sequences from the MATOU database have an average (N50) length of 635 bases and have a mainly eukaryotic association. Protein sequences were inferred from metatranscriptomic data based on features in MATOU and alignments provided through homology searching in MATOU (https://tara-oceans.mio.osupytheas.fr/ocean-gene-atlas/; 08/29/2023). We collected homology hits through homology searches of CBK1, CDC28, KIN1, KIN2 and DFB20 protein sequences from the *S. cerevisiae* strain ATCC 204508 / S288c.

BLASTP was used for homology searches on each of the five representative *S. cerevisiae* protein sequences with the expectation score threshold set to $10^{-10}$. This was due to the large size of the database, but this exact threshold setting was not very consequential for how other more permissive expectation score thresholds (e.g., $10^{-5}$) yielded identical sets of alignment hits. Homology search results included downloadable files for protein alignments, homologs of sequences in a FASTA format, data on genetic sequence variant (i.e., MATOU "Unigene"; distinct transcribed gene sequence) abundance and associated environmental data. The three representative *S. cerevisiae* protein sequences having 500 or more homologous genetic variants identified and classified as being from kingdom fungi were retained for further analysis: CBK1, KIN1 and KIN2.

## 2.3. Data Modeling and Processing

An automated workflow was constructed with R version 4.4.2 (R Core Team, 2019). The files abundance_matrix.csv and environmental_parameters.csv were generated from the Tara Oceans website for each homolog hit search. For the environmental_parameters.csv file, there are 55 columns for environmental data that chart some of the physical and chemical characteristics of each sampling environment (e.g., $NO_2$, carbon total, chlorophyll c3 level, temperature, etc). The OGA_ID (Ocean Gene Atlas identifier; e.g., TARA_038_DCM_0.8-5) encompassed one or more Sample IDs (unique label of the specific sampling instance; e.g., TARA_A100000393) for each sampling location. Sample IDs (row #2 in the original abundance file) were used to join environmental data records (per row names in the environmental_parameters.csv file) to column names in the abundance data file (abundance_matrix.csv).

## 2.4. Partitioning of the Environmental Data

Based on a goal to examine those basic environmental variables known to impact core physiological processes, the environmental variables that were initially selected for study were iron, oxygen, phosphate, salinity, and temperature. For salt, the values for each partition were <35 PSU for low and >36 PSU for high. Temperature values were partitioned as <15°C for low and >25°C for high. Iron values were partitioned as low and high, with low values as <0.006 umol/L, and high values as >0.006 umol/L. Phosphate values were partitioned as <0.25 umol/L for low and >0.5 umol/L for high. Oxygen values for low were <160 umol/kg and >270 umol/kg for high. Frequency counts across these partitioned levels were then tabulated for each MATOU ID (genetic sequence variant). The abundance file was filtered for homologs specific to the fungi kingdom based on files generated from each homology search on the MATOU database linked with taxonomy information in the abundance file. We used these to evaluate alignments of reference sequences to matching genetic sequence variants as based on MATOU IDs indicated from the MATOU database (e.g., MATOU-v1_41411079). Due to having a greater genetic variant abundances, salt, phosphate and iron were retained for further analysis.

## 2.5. Heat Maps of Environmental Conditions and Genetic Sequence Abundance

Heat maps were generated to visualize the abundance of MATOU ID genetic sequence variants that were homologous to each of the cell cycle genes as found across the different environmental conditions evaluated in the study. R packages used for the heat map visualization are pheatmap (version 1.0.12) and dendsort (version 0.3.4). The pheatmap function was used to plot clustered heat maps for the analysis (Kolde, 2019). The dendsort function organized the nodes on the dendrogram showing how MATOU IDs clustered together relative to similarity across environmental conditions (Sakai et al., 2014). Heat maps of environmental condition associations with abundances for different genetic sequences were colorized based on association of the abundance of each MATOU ID genetic sequence variant across each of the environmental condition associations ($x$). Scaling was column-based with respect to the heatmap layout, and based on subtracting from $x$ the mean abundance for each MATOU ID across the full set environmental condition associations (i.e., number of instances for that MATOU ID occurring) divided by the standard deviation.

## 2.6. PERMANOVA

PERMANOVA was used to analyze for significance between the ten environmental conditions and the variants for each of the three cell cycle genes, CBK1, KIN1 and KIN2, with respect to their predictive association with the phyla. The P-value threshold was 0.05 for significance ($P<0.05$). PERMANOVA testing was conducted in RStudio using the adonis2 function in the vegan R package (Dixon, 2009) and was performed using 999 permutations. PERMANOVA tests were done on each of these genes using unscaled abundance counts for each genetic variant (MATOU ID), and the taxonomic association of each genetic variant: Ascomycota, Basidiomycota and other fungi. The Mahalanobis distance method was used based on its utility for examining organismal lineage relationships to environmental variables (De Maesschalck et al., 2000).

## 3. Results

Clustered heat maps were generated to initially visualize how environmental conditions associate with the abundance of MATOU ID genetic sequence variants distinctive for the two most represented fungal phyla, Ascomycota and Basidiomycota, along with association in a category termed as Other Fungi (Figures 1-4). This analysis was performed on each of three different fungal genes associated with cell cycling: CBK1, KIN1 and KIN2. For each gene, the different phylum-level associations for genetic variants show some complexity where, for instance, different genetic variants of different phyla can be found at relatively similar levels across the environmental partitions being measured. In other words, genetic variants of a particular phylum did not all uniquely group together around one set of environmental partition values compared to genetic variants of another phylum.

PERMANOVA was then used to further investigate the significance of environmental association with these levels of abundance. Each of the three genes yielded common outcomes of significance for how abundance levels among these genes' variants among environmental conditions predict phylum-level association (*P<0.05*). For the three fungal genes, the significant association of environmental conditions with phylum-association identified for both the CBK1 and KIN2 genes were intersection between low phosphate and high phosphate levels, and KIN1 did not result in significance (see Tables 1-3).

Table 1. PERMANOVA for phylum-level association of CBK1 genetic sequence variants across environmental conditions ($P<0.05$).

| Environmental Parameter | Df | Sum of Squares | $R^2$ | F | Pr(>F) |
|---|---|---|---|---|---|
| Low Salt | 1 | 2.43 | 0.00666 | 1.2274 | 0.296 |
| High Salt | 1 | 0.17 | 0.00046 | 0.0855 | 0.925 |
| Low Iron | 1 | 2.68 | 0.00737 | 1.3581 | 0.266 |
| High Iron | 1 | 1.12 | 0.00309 | 0.5683 | 0.571 |
| Low Phosphate | 1 | 2.13 | 0.00585 | 1.0771 | 0.338 |
| High Phosphate | 1 | 0.64 | 0.00175 | 0.3223 | 0.732 |
| Low Salt:HighSalt | 1 | 1.52 | 0.00418 | 0.7697 | 0.450 |
| Low Iron:High Iron | 1 | 5.27 | 0.01447 | 2.6656 | 0.065 |
| Low Phosphate:High Phosphate | 1 | 6.12 | 0.01681 | 3.0950 | 0.046* |

Table 2. PERMANOVA for phylum-level association of KIN1 genetic sequence variants across environmental conditions ($P<0.05$).

| Environmental Parameter | Df | Sum of Squares | $R^2$ | F | Pr(>F) |
|---|---|---|---|---|---|
| Low Salt | 1 | 0.612 | 0.00612 | 0.2877 | 0.778 |
| High Salt | 1 | 0.740 | 0.00740 | 0.3478 | 0.659 |
| Low Iron | 1 | 0.421 | 0.00421 | 0.1977 | 0.828 |
| High Iron | 1 | 2.152 | 0.02052 | 1.0113 | 0.360 |
| Low Phosphate | 1 | 3.802 | 0.03802 | 1.7862 | 0.171 |
| High Phosphate | 1 | 2.075 | 0.02075 | 0.9751 | 0.415 |
| Low Salt:HighSalt | 1 | 0.485 | 0.00485 | 0.2277 | 0.806 |
| Low Iron:High Iron | 1 | 0.601 | 0.00601 | 0.2824 | 0.743 |
| Low Phosphate:High Phosphate | 1 | 1.850 | 0.01850 | 0.8693 | 0.419 |

Table 3. PERMANOVA for phylum-level association of KIN2 genetic sequence variants across environmental conditions ($P<0.05$).

| Environmental Parameter | Df | Sum of Squares | $R^2$ | F | Pr(>F) |
|---|---|---|---|---|---|
| Low Salt | 1 | 0.343 | 0.00591 | 0.1950 | 0.823 |
| High Salt | 1 | 2.582 | 0.04452 | 1.4683 | 0.425 |
| Low Iron | 1 | 0.668 | 0.01152 | 0.3800 | 0.729 |
| High Iron | 1 | 0.852 | 0.01468 | 0.4842 | 0.621 |
| Low Phosphate | 1 | 2.830 | 0.04880 | 1.6093 | 0.213 |
| High Phosphate | 1 | 4.295 | 0.07404 | 2.4419 | 0.092 |
| Low Salt:HighSalt | 1 | 1.534 | 0.02645 | 0.8722 | 0.416 |
| Low Iron:High Iron | 1 | 2.330 | 0.04017 | 1.3250 | 0.278 |
| Low Phosphate:High Phosphate | 1 | 7.393 | 0.12747 | 4.2040 | 0.020* |

## 4. Discussion

Visual analysis indicates some interdependence of genetic variant sequences with environmental conditions for each of the three cell cycle genes, and there was contrasted clustering between high versus low partitionings of environmental conditions for iron, salt and phosphate. This was found to be statistically significant for both CBK1 and KIN2 genes for how phylum-level associations of genetic sequence variant abundance occur for low phosphate versus high phosphate conditions. Although KIN1 and KIN2 are similar genes, they have previously been reported to have some differential response to stress in *Arabidopsis thalania* (Kurkela & Borg-Franck, 1992).

Determining the chemical nature and concentrations of dissolved substances may be challenging due to the range of low to high concentrations found for different ions, and even more prevalent ions like sodium and potassium are difficult to determine accurately. Differentiating between substances with similar chemical properties, such as phosphate, arsenic, calcium, strontium, chloride, bromide, and iodide, has also been reported as being problematic (Uçak & Aydın, 2022; Sverdrup, 1946). The physical and chemical environment is not constant. Salinity, for example, changes depending on precipitation, erosion, and sunlight. Salinity can range from 5% near some coasts to 45% in saltier waters. An example of variation between larger areas of oceanic waters is with the Black Sea being at about 2% salinity while the Red Sea is at about 4.5% salinity (Keener-Chavis & Sautter, 2002).

In general, the level of fluctuation associated with physical and chemical conditions of oceanic waters at specific locations would by itself be an essential area for further study. Further application of our method would potentially involve prospective studies increasing the depth of metatranscriptomic sequencing for genetic variant identification and studies of both fungal and non-fungal microorganisms. Greater monitoring of associated environmental conditions across time within the ecosystem being studied would help to evaluate further the association between genetic variation and local ranges of environmental conditions. These approaches, along with a greater amount of data sampling to ensure that enough data are collected, would support a robust, accurate finding. Ensuring that, for instance, a threshold of abundance is identified for each represented lineage and set of environmental conditions along with further potential treatment of zero counts are active areas of current research into what may be expected from PERMANOVA-based studies of microbiomes (Clarke et al., 2006). Further progression of this research could overall provide unique and relevant insights on how adaptive change inferred from genetic variation may relate to altered cellular pathway functionality. The methodological and analytical framework of this investigation may go beyond analyses of environmental variation in the ocean, and provide insight on other similar instances of altered cellular pathway functionality or phenotype in living organisms in other environments where there is some orthologous genetic composition across these organisms.

## 4. Conclusion

This work overall demonstrates how to analyze a large complex set of assorted and sporadic data on genetic variation and associated environmental conditions. This work further shows how levels of phosphate in ocean waters may control for different types of oceanic fungi with varying genetic backgrounds.

### Acknowledgements

This research was supported by an appointment to the Intelligence Community Postdoctoral Research Fellowship Program at North Carolina Agricultural and Technical State University, administered by Oak Ridge Institute for Science and Education through an interagency agreement between the U.S. Department of Energy and the Office of the Director of National Intelligence. This research was also supported by the Chancellors Title III PhD Fellowship at North Carolina Agricultural and Technical State University.

### Authors' Contributions

SAD utilized and collected data from the NCBI database, the Tara Oceans databases, and the Yeast KPI database. SAD and SHH conceived of the study, jointly developed R code for data integration and computation, and were both major contributors to the writing and editing of the manuscript.

### Competing Interests

The authors declare no competing interests.

# References

Braun, A. (1856) - Ueber Chytridium, eine Gattung einzelliger Schmarotzergewachse auf Algen und Infusorien. *Abhandl. Berlin. Akad.* 1855: 21-83

Breitkreutz, A., Choi, H., Sharom, J. R., Boucher, L., Neduva, V., Larsen, B., Lin, Z.-Y., Breitkreutz, B.-J., Stark, C., Liu, G., Ahn, J., Dewar-Darch, D., Reguly, T., Tang, X., Almeida, R., Qin, Z. S., Pawson, T., Gingras, A.-C., Nesvizhskii, A. I., & Tyers, M. (2010). A Global Protein Kinase and Phosphatase Interaction Network in Yeast. *Science*, vol. 328, no. 5981, pp. 1043–1046. https://doi.org/10.1126/science.1176495

Clarke, K. R., Somerfield, P. J., & Chapman, M. G. (2006). On resemblance measures for ecological studies, including taxonomic dissimilarities and a zero-adjusted Bray–Curtis coefficient for denuded assemblages. *Journal of Experimental Marine Biology and Ecology*, *330*(1), 55-80. https://doi.org/10.1016/j.jembe.2005.12.017

De Maesschalck, R., Jouan-Rimbaud, D., & Massart, D. L. (2000). The Mahalanobis distance. *Chemometrics and Intelligent Laboratory Systems*, vol. 50, no. 1, pp. 1–18, doi: https://doi.org/10.1016/S0169-7439(99)00047-7

Dixon, P. (2003). VEGAN, a package of R functions for community ecology. *Journal of Vegetation Science*, vol. 14, no. 6, pp. 927–930, doi: https://doi.org/10.1111/j.1654-1103.2003.tb02228.x

Fell, Jack W.. "6 Yeasts in marine environments". *Marine Fungi*, edited by E. B. Gareth Jones and Ka-Lai Pang, Berlin, Boston: De Gruyter, 2012, pp. 91-102. https://doi.org/10.1515/9783110264067.91

Geer, L. Y., Marchler-Bauer, A., Geer, R. C., Han, L. , He, J., et al. (2010). The NCBI biosystems database. *Nucleic Acids Research*, *38*(suppl_1), D492-D496. https://doi.org/10.1093/nar/gkp858

Hagler, A. (1987). Ecology of aquatic yeasts. *The Yeasts*. *1*, 181-205.

Hingamp, P. (2018). The Ocean Gene Atlas: exploring the biogeography of plankton genes online. *Nucleic Acids Research, 46(W1), W289-W295*. https://doi.org/10.1038/ismej.2013.59

Morris, E. O. (1968). Yeasts of marine origin. *Oceanogr Mar Biol Ann Rev*. *6*, 201-230.

Keener-Chavis, P. & Sautter, L. Of sand and sea: teachings from the southeastern shoreline. Accessed: Feb. 16, 2023. [Online]. Available: https://repository.library.noaa.gov/view/noaa/37694/noaa_37694_DS1.pdf

Kolde, R. (2019). pheatmap: Pretty Heatmaps. *R-Packages*, Jan. 04, 2019. https://cran.r-project.org/package=pheatmap

Kriss, A. E. (1967). *Microbial population of oceans and seas* (No. QR106 K7213 1967).

Kurkela, S., Borg-Franck, M. Structure and expression of *kin2*, one of two cold- and ABA-induced genes of *Arabidopsis thaliana*. *Plant Molecular Biology* 19, 689–692 (1992). https://doi.org/10.1007/BF00026794

Kutty, S. N. & R. Philip, R. (2008). Marine Yeasts-a review. *Yeast*. 24:465-483. https://doi.org/10.1002/yea.1599

R Core Team (2019). R: The R Project for Statistical Computing. *R-project.org*, https://www.r-project.org

Sakai, R., Winand, R., Verbeiren, T., Moor, A. V., & Aerts, J. (2014). dendsort: modular leaf ordering methods for dendrogram representations in R. *F1000Research*, vol. 3, p. 177, Jul. 2014. https://doi.org/10.12688/f1000research.4784.1

Sverdrup, H. U., Johnson, M. W, & Fleming, R. H. (1946). *The oceans: their physics, chemistry, and general biology*. New York: Prentice-Hall. http://ark.cdlib.org/ark:/13030/kt167nb66r/

Uçak, Ş. Ş. & Aydın, A. A novel thiourea derivative for preconcentration of copper(II), nickel(II), cadmium(II), lead(II) and iron(II) from seawater samples for Flame Atomic Absorption Spectrophotometry. *Marine Pollution Bulletin*, vol. 180, p. 113787, Jul. 2022, doi: https://doi.org/10.1016/j.marpolbul.2022.113787

Villar, E., Vannier, T., Vernette, C., Lescot, M., Cuenca, M., Alexandre, A., Bachelerie, P., Rosnet, T., Pelletier, E., Sunagawa, S. & Hingamp, P. (2018). The Ocean Gene Atlas: exploring the biogeography of plankton genes online. *Nucleic Acids Research, 46(W1), W289-W295*. https://doi.org/10.1093/nar/gky376

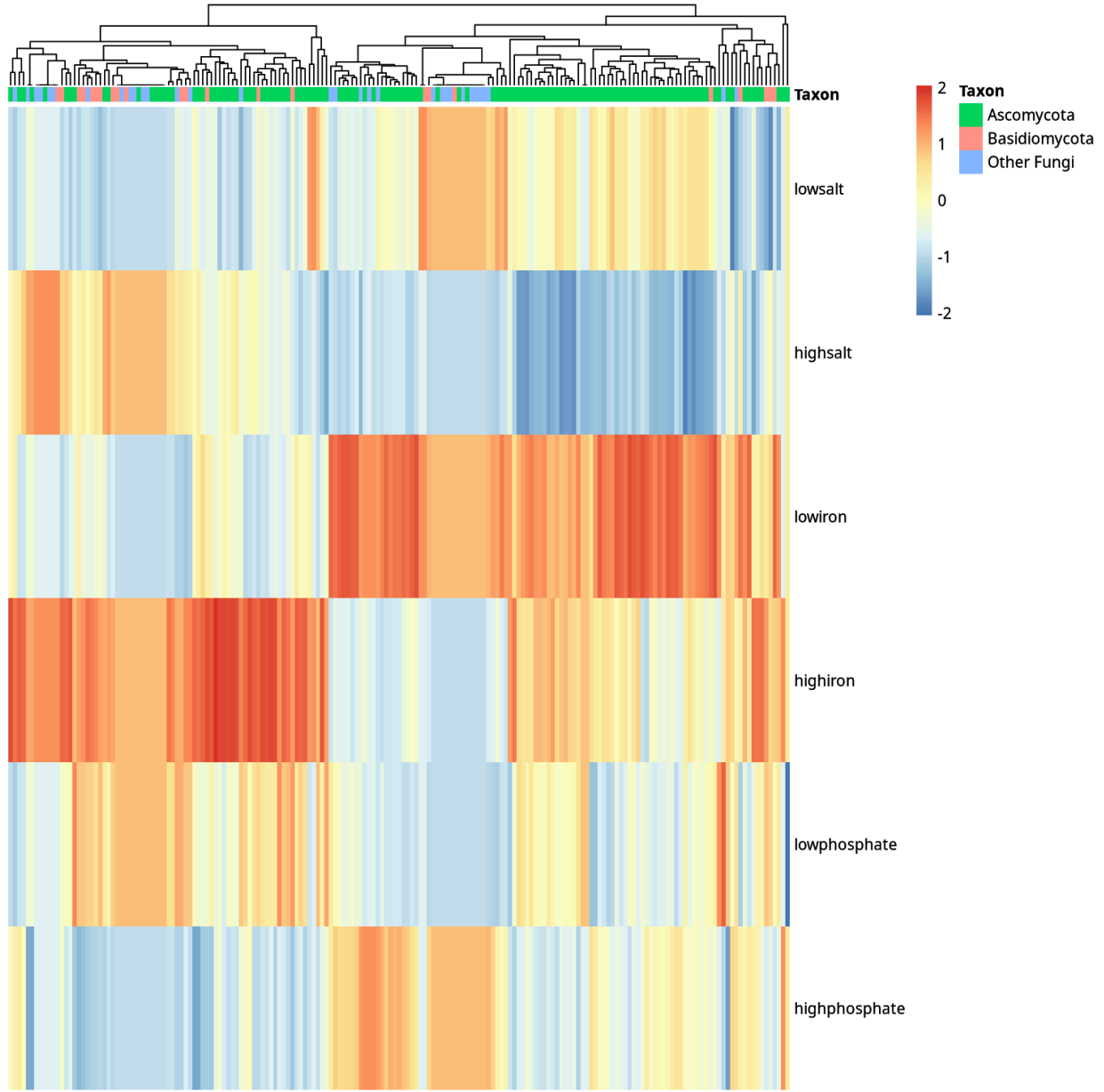

**Figure 1.** Clustered heat map for CBK1 genetic sequence variant abundances across six environmental conditions. Relative abundance is scaled based on the number of standard deviations away from the mean abundance across the six environmental conditions for each CBK1 genetic sequence variant. Phylum-level associations are also shown.

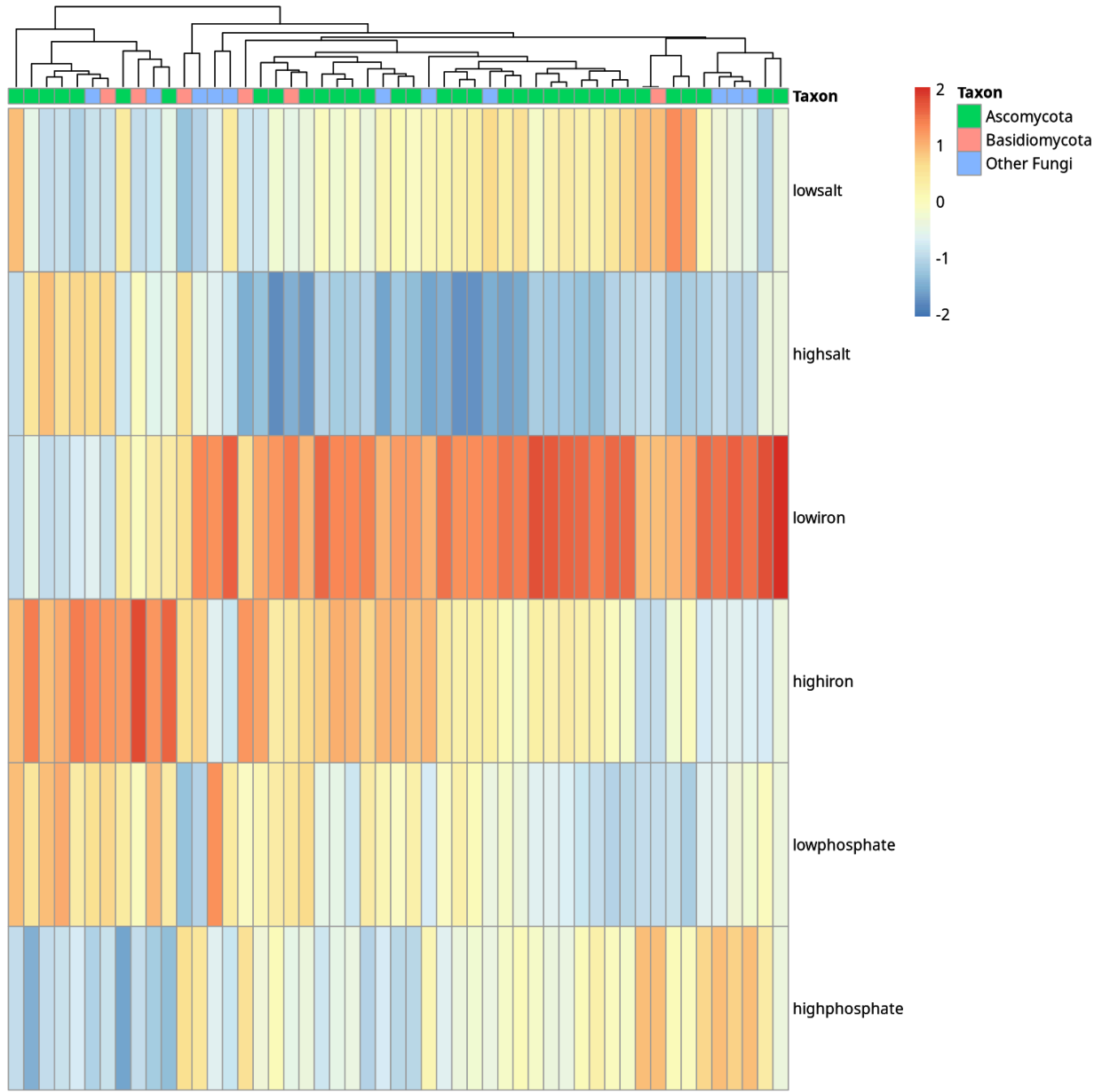

**Figure 2.** Clustered heat map for KIN1 genetic sequence variant abundances across six environmental conditions. Relative abundance is scaled based on the number of standard deviations away from the mean abundance across the six environmental conditions for each KIN1 genetic sequence variant. Phylum-level associations are also shown.

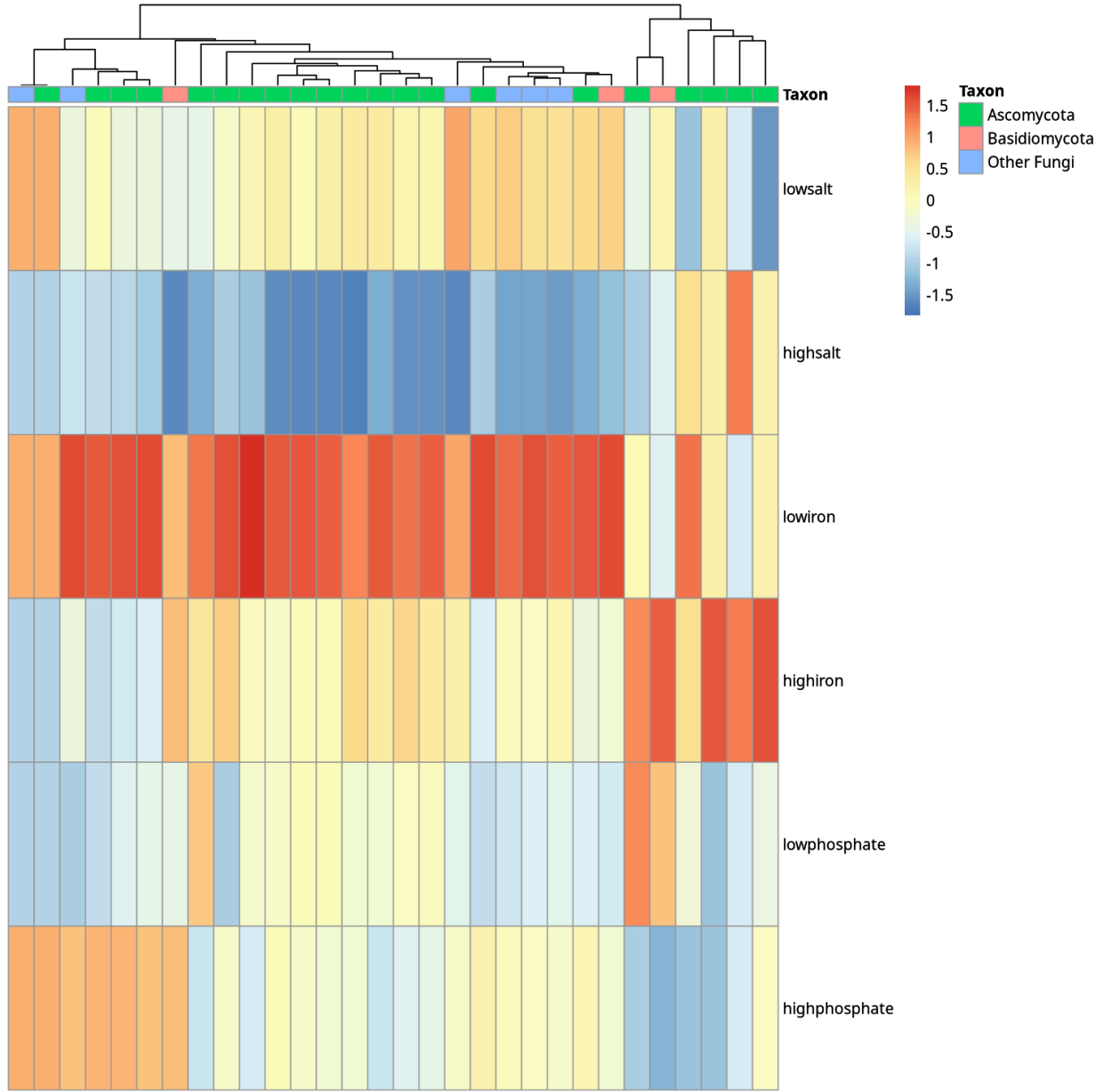

**Figure 3.** Clustered heat map for KIN2 genetic sequence variant abundances across six environmental conditions. Relative abundance is scaled based on the number of standard deviations away from the mean abundance across the six environmental conditions for each KIN2 genetic sequence variant. Phylum-level associations are also shown.

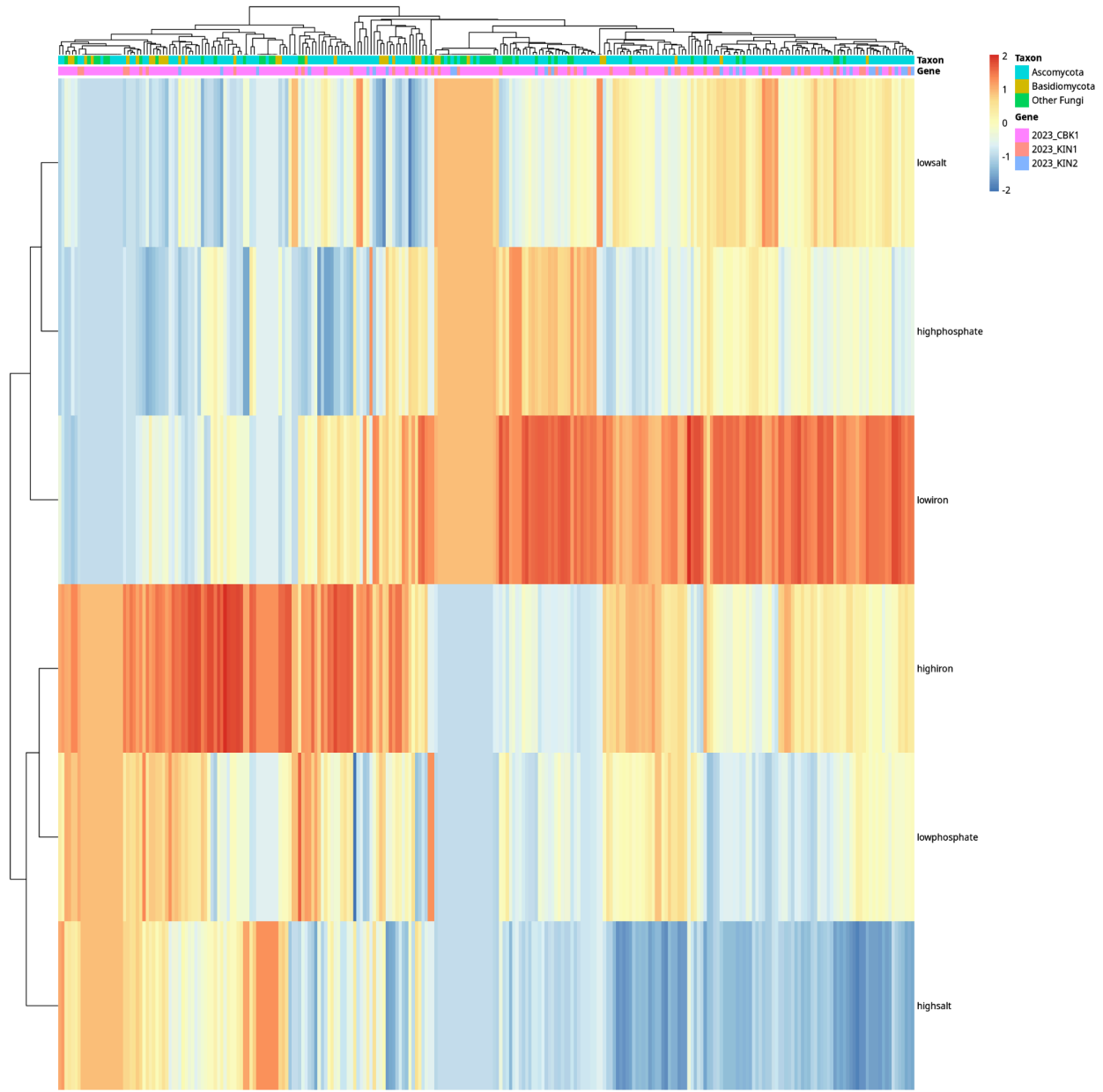

**Figure 4.** Clustered heat map for genetic sequence variants of the three genes CBK1, KIN1 and KIN2 based on abundance across six environmental conditions. Relative abundance is scaled based on the number of standard deviations away from the mean abundance across the six environmental conditions for each specific genetic sequence variant. Phylum-level associations are also shown.